\documentclass[modern]{aastex63}

\shorttitle{Dynamics of SN Formed Dust}
\shortauthors{Slavin et al.}
\graphicspath{{./}{figures/}}
\begin{document}

% mm [messed with the punctuation]
\title{The Dynamics, Destruction, and Survival of Supernova-Formed Dust Grains}

\correspondingauthor{Jonathan D. Slavin}
\email{jslavin@cfa.harvard.edu}

\author[0000-0002-7597-6935]{Jonathan D. Slavin}
\affiliation{Center for Astrophysics $\vert$ Harvard \& Smithsonian,
60 Garden Street,
Cambridge, MA 02138, USA}

\author[0000-0001-8033-1181]{Eli Dwek}
\affiliation{Observational Cosmology Lab, NASA Goddard Space Flight Center,
Mail Code 665, Greenbelt, MD 20771, USA}

\author[0000-0003-0064-4060]{Mordecai-Mark Mac Low}
\affiliation{American Museum of Natural History, 79th Street at Central Park
West, New York, NY 10024, USA}
\affiliation{Center for Computational Astrophysics, Flatiron Institute, New
York, NY, 10010, USA}

\author[0000-0001-7301-5666]{Alex S. Hill}
\affiliation{Space Science Institute, Boulder, CO 80301, USA}
\affiliation{Department of Computer Science, Math, Physics, and Statistics,
University of British Columbia, Okanagan Campus, 3187 University Way, Kelowna,
BC V1V 1V7 Canada}
\affiliation{Dominion Radio Astrophysical Observatory, Herzberg Program in
Astronomy and Astrophysics, National Research Council Canada, Penticton, BC
V2A 6J9 Canada}

\begin{abstract}
Observations have demonstrated that supernovae efficiently produce dust. This
is consistent with the hypothesis that supernovae and asymptotic giant branch
stars are the primary producers of dust in the Universe. However, there has
been a longstanding question of how much of the dust detected in the interiors
of young supernova remnants can escape into the interstellar medium. We
present new hydrodynamical calculations of the evolution of dust grains that
were formed in dense ejecta clumps within a Cas~A-like remnant. We follow the
dynamics of the grains as they decouple from the gas after their clump is hit
by the reverse shock. They are subsequently subject to destruction by thermal
and kinetic sputtering as they traverse the remnant.  Grains that are large
enough ($\sim 0.25\,\mu$m for silicates and $\sim 0.1\,\mu$m for carbonaceous
grains) escape into the interstellar medium while smaller grains get trapped
and destroyed. However, grains that reach the interstellar medium still have
high velocities, and are subject to further destruction as they are slowed
down. We find that for initial grain size distributions that include large
($\sim 0.25 - 0.5\,\mu$m) grains, 10--20\% of silicate grains can survive,
while 30--50\% of carbonaceous grains survive even when the initial size
distribution cuts off at smaller ($0.25\,\mu$m) sizes. For a 19 M$_{\odot}$
star similar to the progenitor of Cas~A, up to 0.1 M$_{\sun}$ of dust can
survive if the dust grains formed are large. Thus we show that supernovae
under the right conditions can be significant sources of interstellar dust.
\end{abstract}



\section{Introduction} \label{sec:intro}

% mm: I think we're going to need to add references throughout these introductory paragraphs.
Dust plays a variety of important roles in the interstellar medium (ISM) in a
wide array of environments. In low-density warm and cold neutral regions it
can be an important source of heat via photoelectric emission \citep[e.g.][and
references therein]{Wolfire_etal_2003}. In molecular clouds dust is important
as a site for molecule formation \citep[e.g.][]{Hollenbach+Salpeter_1971}, as
well as shielding molecules from ultraviolet radiation, and cooling the gas at
the highest densities. Dust can also be an important coolant for hot gas via
conversion of thermal energy to infrared (IR) emission \citep{Dwek_1987}. As a
repository of metals, dust regulates the gas phase abundances and metal
transport in galaxies. 

The evolution of dust is complex and is driven by many of the key processes
involved in galactic evolution. The cores of dust grains are known to form in
the dense ejecta of supernovae (SNe) and in the slow, dense winds from evolved
stars, especially asymptotic giant branch stars. Interstellar grains are
destroyed primarily by sputtering, which is enhanced behind shock waves
because of the presence of high pressure hot gas or, in radiative shocks, high
compression, which leads to betatron acceleration and accompanying inertial
sputtering and grain-grain collisions (which shatter grains, changing the size
distribution). Thus grain destruction is tightly coupled to the rate of energy
injection into the ISM by SNe. The persistent problem of the apparent
imbalance of grain destruction rates and creation rates \citep[see
e.g.][]{Jones_etal_1994} has lead to the idea that accretion of gas onto
pre-existing grain cores is important to maintaining the levels of gas phase
depletion inferred for the ISM
\citep{Draine+Salpeter_1979b,Dwek+Scalo_1980,Draine_2009}. While such
accretion could be the solution to the grain destruction problem, it is
nonetheless important that grain cores are injected at an adequate rate since
it does not appear to be possible to create such cores in the ISM.

SNe are known creators of dust as has been shown directly by infrared (IR)
observations
\citep{Barlow_etal_2010,Gomez_etal_2012,Arendt_etal_2014,Matsuura_etal_2015}. 
The most notable dust sources are SN 1987A, Cas A, The Crab Nebula and
G54.1+0.3. A more complete list of Galactic and extragalactic sources can be
found in \citet{Sarangi_etal_2018}. In particular far-IR observations, for
example with Herschel, have shown that remnants such as Cas A and SN 1987A
have created $\sim 1$ M$_\sun$ of dust in the densest regions of their ejecta.
This dust, however, is cold and has not yet encountered the reverse shock that
will eventually heat the interior of the remnants to very high temperatures
and may destroy significant amounts of the newly created dust
\citep{Dwek_2005}. Though it is known that some dust created in SNe does
escape, e.g.\ based on the isotopes found in some pre-solar grains in
meteorites \citep[e.g.][]{Hoppe_etal_2015}, whether or not SNe are important
sources of interstellar dust in the present-day universe remains uncertain.
There is also keen interest in whether SNe in high-redshift galaxies can
produce the high levels of dust observed in them
\citep{Bertoldi_etal_2003,Gall_etal_2011a,Dwek_2011a}, since in many cases the
galaxies are too young for stars to have evolved to the AGB phase.

There have been several studies of the question of the survival of dust formed
in SNe
\citep{Dwek_2005,Nozawa_etal_2007,Bianchi+Schneider_2007,Nath_etal_2008,Silvia_etal_2012,Silvia_etal_2010,Biscaro+Cherchneff_2016,Micelotta_etal_2016,Bocchio_etal_2016,Kirchschlager_etal_2019},
mainly focused on how much destruction occurs when the reverse shock sweeps
over the newly formed grains. These studies have used a range of methods and
made predictions ranging from very low destruction to complete destruction of
the dust.  Of these previous studies, the ones that have employed
hydrodynamical calculations
\citep{Silvia_etal_2010,Silvia_etal_2012,Kirchschlager_etal_2019} have all
used a plane parallel shock encountering a cloud, or ejecta clump. Those
previous works have examined how much grain destruction occurs due to
sputtering when the clump is shocked and heated. \citet{Silvia_etal_2012}
assumed that the grains were tightly coupled to the gas and that thermal
sputtering was the dominant process. \citet{Micelotta_etal_2016} took a
different approach, examining how grains in the clumps get sputtered when they
decouple from the gas and stream out of the clump as it is decelerated by the
reverse shock. \citet{Kirchschlager_etal_2019} included more processes,
including inertial sputtering, grain-grain collisions and decoupling of the
gas and dust. They treat the dust in post-processing using what they term a
dusty-grid approach. This approach amounts to a multi-fluid treatment because,
though the dust can move relative to the gas, grains from one grid zone cannot
interpenetrate those from a different grid zone and grains from each zone all
move together.  We discuss these assumptions further below.  These studies are
important since they have demonstrated that substantial grain destruction can
occur when the reverse shock hits an ejecta clump. However, grains that
survive the reverse shock are still within the hot gas of the evolving SN
remnant and thus subject to continued sputtering. 

Determining if grains escape the SN remnant requires tracking the motion of
the grains relative to the shock front in the evolving remnant. In this paper
we present results from new hydrodynamical calculations in which we track the
grains as they evolve with the remnant. We simulate a remnant that is similar
to Cas A including very small and high density ejecta clumps. We start our
simulations with grains inside the clumps and follow the dust trajectories as
they are sputtered because of their interaction with the gas and suffer drag
when they have substantial velocity relative to the surrounding gas. We follow
the grains until they either escape the remnant or are destroyed. Escaping
grains are further destroyed while being slowed in the ISM. Our results are
important for assessing under which conditions grains can fully escape the SN
remnant to enter the general ISM and thus the rate at which SNe can supply
dust to the ISM.

The survival of the dust depends on the morphology of the SN ejecta, and that
of the ambient medium of the progenitor star. To simulate a realistic scenario
we have created models that use the Cassiopeia A SN remnant as an example.
That is to say, we use the observations of Cas A to guide our parameter
choices while not attempting to match the remnant in detail. Thus, for
example, we use clumpy ejecta that are roughly of the size determined for
those in Cas A. On the other hand we make no attempt to model the jet or the
complex abundance variations that are seen in Cas A. Our goal is to focus on
aspects of the remnant that have the most impact on the evolution of the dust
generated within the remnant. Our overall aim has been to create models that
may be widely applicable to core-collapse SN remnants that produce dust.

\section{Methods} \label{sec:methods}
\subsection{Hydrodynamical Simulations}
For the hydrodynamical simulations presented here we have used the publicly
available and well-tested FLASH
code\footnote{\url{http://flash.uchicago.edu/site/flashcode/}}
\citep{Fryxell_etal_2000,Dubey_etal_2012}.
Because of the need for very high dynamic range in order to resolve the ejecta
clumps, $R_\mathrm{clump} \sim 10^{16}$ cm \citep{Fesen_etal_2011}, while
following the remnant evolution to relatively large size, $\sim 20$ pc, we
have run our simulations in two dimensions (2D) with cylindrical symmetry. We
have made use of FLASH's adaptive mesh refinement capabilities with eight
levels of refinement, which leads to a resolution at full refinement of
$7.53\times10^{15}$ cm ($0.00244$ pc). For our grid, which extends from $0$ to
$20$ pc in $r$ and $-20$ to $+20$ pc in $z$, this would be, if fully refined,
$8192 \times 16384$ zones.

Dust grains in the ISM are coupled to the gas via drag forces and the magnetic
field. We have not included the magnetic field in our calculations for reasons
that we discuss below. We have included both drag and sputtering (grain
erosion) and have allowed for the independent motion of individual grains. The
FLASH code provides units for following both active and passive particles
where the former act on the gas (by gravitational force in the provided unit)
and are free to move independently of the gas while the latter simply move
with the gas acting as Lagrangian tracers. Evolution of the particle mass is
not included in the active particle unit.  Dust does not fit neatly into
either category since typically it does not include enough mass to
significantly affect the gas dynamics and yet couples only imperfectly to the
gas via drag (and the magnetic field) and can thus move independently of the
gas. For this reason we needed to develop our own unit that builds on the
existing active particle unit. We also developed a short range force unit that
implements gas drag as well as sputtering. 

Another modification to the basic code base of FLASH that we have made is to
add radiative cooling via a lookup table. Our approach has been similar to
that in the user-contributed supplemental SutherlandDopita
\citep{Sutherland+Dopita_1993} unit (under {\tt physics/sourceTerms/Cool})
though it differs in detail. The table of cooling coefficients we have used
was generated by us using the code Cloudy. We ran Cloudy \citep[version
17.00]{Ferland_etal_2017} as a subroutine using abundances like those inferred
for the O-rich ejecta in Cas A (essentially pure O) over a large range of
temperatures. This cooling curve would not be appropriate for the smooth
ejecta nor for the circumstellar medium or ISM around the remnant. However,
this is not a problem for our simulations, since cooling is unimportant in
those regions over the time span of our simulations because of their low
densities.

\subsection{Initial Conditions}
\begin{figure}[ht!]
    \centering
    \includegraphics[width=0.6\textwidth]{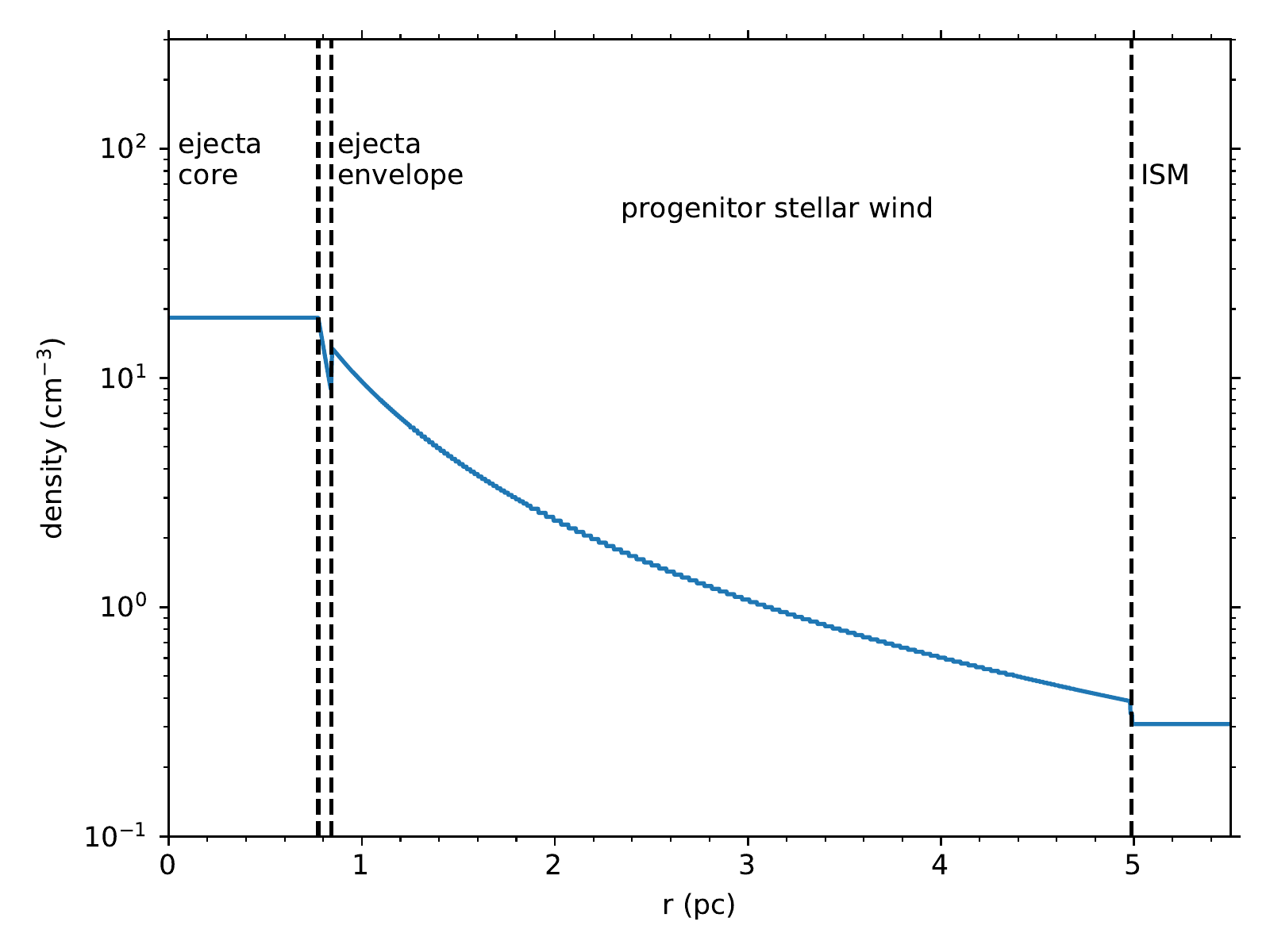}
    \caption{Initial density profile for our simulations. In this 1D rendering
    the dense ejecta clumps are not included. Note that here $r$ is the radial
    distance from the explosion center, not the cylindrical distance from the
    $z$ axis.}
    \label{fig:init_dens}
\end{figure}

We initiate our simulations with fast-expanding ejecta that contain embedded
dense clumps. The ejecta initially have a constant density core with density
of the smooth ejecta of $\rho_{sm} = 3.831\times10^{-23}\,$g cm$^{-3}$.
Outside of the core is the envelope in which the density declines steeply,
$\rho \propto r^{-9}$. (Note that here and in our discussion below regarding
the progenitor stellar wind region, $r$ is the spherical distance from the
origin, not the cylindrical distance from the symmetry axis.) The clumps are a
factor of 100 times denser than the surrounding smooth ejecta with initial
radii of $3\times10^{16}$ cm. This is somewhat larger than observed, though
the observations are of clumps that have been shocked and compressed. The
clumps are in pressure equilibrium with the ejecta, though this does not
matter much since the thermal energy is very small compared to the kinetic
energy. They are limited to the smooth ejecta region initially and their
number is set by our assumed volume filling factor for the clumps of 2\%. For
our assumed parameters this leads to 61 clumps, which are scattered randomly
within the ejecta core and are sharp edged circles (tori in cylindrical
symmetry). The mean density in the core is $\langle \rho_\mathrm{core} \rangle
= ((1 - f_{cl}) + \chi f_{cl}) \rho_{sm}$ where $f_{cl}$ is the clump filling
factor, $\chi$ is the ratio of clump density to smooth density (100 in our
calculations) and $\rho_{sm}$ is the density of the smooth ejecta. For our
assumptions, $\langle \rho_\mathrm{core} \rangle = 1.141\times10^{-22}$ g
cm$^{-3}$. For the purposes of calculating the cooling in the clumps, they are
assumed to be essentially pure oxygen (see discussion below). The initial
velocity profile is assumed to be linear, $v \propto r$, from the origin to
the edge of the ejecta envelope.  Our assumed initial density profile is shown
in figure \ref{fig:init_dens}, though without the dense clumps.

The combination of our assumed ejecta mass, 3.5 M$_\sun$, explosion energy,
$1.5\times10^{51}$ ergs, and maximum speed, 9000 km s$^{-1}$, along with the
mean density in the ejecta core, fix the size of the core relative to outer
edge of the ejecta envelope. In our case that ratio is 0.919. With our assumed
ejecta radius of 0.85 pc, this makes the radius of the core 0.78 pc. The dust
grain particles in the simulations are all started randomly scattered inside
the dense clumps. We have used 40 grains per clump in results presented here.
We have carried out separate runs for each grain size and type discussed. All
grains start with the same size for a given run.

Outside of the ejecta envelope we assume a circumstellar medium density
declining as $\rho \propto r^{-2}$, under the assumption that the explosion
occurred in the stellar wind-blown bubble of the progenitor. There are
observational constraints on the density of the circumstellar medium at the
current location of the forward shock, $R_b \approx 2.5$ pc
\citep{Willingale_etal_2003,Lee_etal_2014}, but the constraints are not tight.
We found that the value of 2.07 cm$^{-3}$ in \citet{Micelotta_etal_2016} was
too high to allow for the forward shock to reach $R_b \approx 2.5$ pc by $\sim
330$ yr. In addition, the total mass in the circumstellar bubble would be very
high unless the bubble were quite small. We can connect the parameters of the
circumstellar bubble with the stellar parameters and observed characteristics
of the remnant by using a combination of the density at the (current) forward
shock, the progenitor mass and the ejecta mass. For this we have assumed that
the progenitor of Cas A was a 19 M$_\sun$ star, that the mass of shocked
circumstellar medium inside of $R_b$ is 6.2 M$_\sun$, and the ejecta mass was
3.5 M$_\sun$. This leads to a density at 2.5 pc of 1.5 cm$^{-3}$, which is
roughly consistent with \citet{Lee_etal_2014}, who found a density at the
forward shock of $\sim 1$~cm$^{-3}$.  With our choice of parameters, the
bubble extends to 5 pc, outside of which we assumed a constant density ISM
with $n = 0.309$ cm$^{-3}$.

\subsection{Extrapolating Grain Evolution}\label{sect:extrap}
In Section~\ref{sec:results}, we present simulation results for a time of
8000~yr.  While that allowed for grains to escape
ahead of the SN remnant shock in some cases, in other cases the grains
remain inside the shock. In either case the question of the final
state of a grain when its speed finally drops below the sputtering
threshold in the ambient ISM remains undetermined. Thus we have found it
necessary to extrapolate the evolution of the grains.

The extrapolation that we have performed has two parts. First, we used the
radially averaged profiles of pressure, density and radial velocity from the
final time step of the 2D simulation to initiate a one-dimensional (1D;
spherically symmetric) simulation that we ran on a large (50 pc) grid for
$10^5$ yr. That allowed us to calculate the shock radius as a function of
time. With the assumption that the fluid variables follow a Sedov-Taylor type
similarity solution \citep[see e.g.][]{Cox+Franco_1981}, we can use the time
evolution of the shock radius to find values of density, temperature, and gas
velocity for locations inside the remnant for late times. We have compared
such profiles with the 1D simulation profiles and found the match very good in
the outer regions of the remnant where the dust is concentrated at late times.
For the 1D simulation, we included radiative cooling with a cooling curve
appropriate for solar abundances. With this cooling, the remnant only starts
to go radiative at about $7\times10^4$ yr. After this point the Sedov-Taylor
profile no longer provides a good approximation to the radial profiles of the
fluid variables and we stop the extrapolation. 

We used the 1D extrapolated remnant evolution in conjunction with the
equations for drag on the grains and thermo-kinetic sputtering (see
\S\ref{sect:processing}) to calculate the motion and mass evolution of the
grains beyond the 8000 yr endpoint of our 2D simulations. In fact for runs
with large initial grain sizes, some of the grains leave the grid before the
end of the simulation. For those grains we use their last position and speed
to extrapolate their evolution. In general the vast majority of the grains
have either been effectively completely destroyed ($m/m(0) < 0.005$) or have
fully escaped the remnant such that they are beyond the shock and have speeds
that exceed the shock speed when we stop the extrapolation at $t =
7\times10^4$ yr. There are typically a small number of grains that have not
fully escaped the remnant but appear likely to do so. These are grains that
started close to the explosion site such that their initial velocity was
relatively low. Some complex behaviors are seen, such as grains that get ahead
of the shock but are then caught by the shock after getting slowed by drag in
the ISM. Also, even grains that have fully escaped the remnant, cannot be
considered to be part of the ISM until they slow to below the sputtering
threshold speed, which we calculate to be $v_\mathrm{thresh} \approx 20$ km
s$^{-1}$. For that reason we further extrapolate the grain evolution past the
cooling time of the 1D simulation at $7\times10^4$ yr, using the results as
plotted in Figure \ref{fig:mm0_v} to calculate the further reduction of mass
that will result as the grains are slowed below $v_\mathrm{thresh}$. 

\subsection{Grain Processing}\label{sect:processing}
Grain sputtering is caused by gas atoms and ions colliding with dust grains
with enough energy to knock atoms off the grain. When the relative speed of
the grain to the gas is much less than the thermal speed of the gas particles
(typically hot gas, $T \gtrsim 10^6\,$ K) then sputtering is primarily thermal
since the velocity distribution of incident particles is Maxwellian. If the
grain speed is large compared to the thermal speeds (typically cold gas, $T
\lesssim 10^4\,$ K), the sputtering is referred to as inertial (or kinetic),
since the velocity distribution of the incident particles is determined by the
relative velocity of the grains through the gas. In the general case, both the
thermal speed of the particles and the gas-grain velocity are important. In
that case the sputtering rate depends on the yield integrated over a
Maxwellian particle distribution skewed by the gas-grain relative velocity and
so depends on both the gas temperature and the relative velocity. Thus the
sputtering rate is 
\begin{equation}\label{eqn:dmdt}
    \frac{dm_{gr}}{dt} = -\pi a_{gr}^2 m_{sp} n_\mathrm{H}\sum_{i} A_i \langle
    Y(E_i) v\rangle
\end{equation}
where $\langle Y v\rangle$ is the yield times the relative gas-grain speed
integrated over the velocity distribution function of the gas particles in the
frame of the grain (typically a skewed Maxwellian), $m_{sp}$ is the average
mass of a sputtered atom, $n_\mathrm{H}$ is the number density of H in the
gas, $A_i$ is the gas phase abundance of gas particle $i$ and the sum is over
incident gas particle types (i.e. elements).  The skewed Maxwellian speed
distribution can be expressed as
\begin{equation}
    f(v,v_{gr},T) = \sqrt{\frac{m}{2\pi k
    T}}\left(\frac{v}{v_{gr}}\right)\left(\exp\left(-\frac{m(v -
    v_{gr})^2}{2kT}\right) - \exp\left(-\frac{m(v +
    v_{gr})^2}{2kT}\right)\right)
\end{equation}
when the integration is over speed \citep[see, e.g.][]{Dwek+Arendt_1992}. Note
that $Y(E_i)$ above is the angle integrated value rather than the normal
incidence value for the yield. We have used the yields from
\citet{Nozawa_etal_2006}, which are based on \citet{Tielens_etal_1994},
adopting the values for MgSiO$_3$ for the silicate grains and those for
amorphous C for the carbonaceous grains. To ease the computational burden that
would be required to calculate the integral for each grain and time step, we
have created tables of $\langle Y v\rangle$ for a large range of temperatures
and grain speeds relative to the gas and for each grain type (carbonaceous and
silicate) and incident atom type (including H, He, C and O). The tables are
read in during initialization and values of $\langle Y v \rangle$ are found
through interpolation to calculate the sputtering rate during the simulations.

We include the effects of drag on the grains as well, which is essential to
calculating their evolution. The grain velocity evolves as
\begin{equation}\label{eqn:dvdt}
    \frac{d\mathbf{v}_{gr}}{dt} = \frac{-\pi a_{gr}^2
    \rho_\mathrm{gas}}{m_{gr}} \xi (\mathbf{v}_{gr} -
    \mathbf{v}_\mathrm{gas})|\mathbf{v}_{gr} - \mathbf{v}_\mathrm{gas}|,
\end{equation}
where $a_{gr}$ is the grain radius, $m_{gr}$ is the grain mass,
$\mathbf{v}_{gr}$ is the grain velocity, $\mathbf{v}_\mathrm{gas}$ is the gas
velocity and
\begin{equation}
    \xi = \left(1 + \frac{128}{9\pi}\frac{k T}{m v^2}\right)^{1/2}
\end{equation}
\citep{Baines_etal_1965,Draine+Salpeter_1979a}. Here $v$ is the absolute value
of the relative gas-grain speed and $T$ is the gas temperature. We note that
we have ignored the plasma drag. As we discuss further below, we have found it
is small compared with the direct drag for the conditions in our simulations.
Ignoring the plasma drag relieves us of the need to calculate the grain
charge, which would involve significant additional computational overhead.

\subsubsection{Grain Slowing in the ISM}
As discussed above, to examine the remnant evolution over longer timescales,
we used our 2D simulation results, radially averaged, to initiate a 1D
simulation in a much larger grid (50 pc) for a time of $10^5$ yr. From that
simulation we derive the long term evolution of the shock size and speed in
the, assumed uniform, ISM. However even at the end of the 1D simulation, most
grains have not slowed enough in the ISM to have stopped sputtering.

\begin{figure}[ht!]
    \centering
    \includegraphics[width=0.7\textwidth]{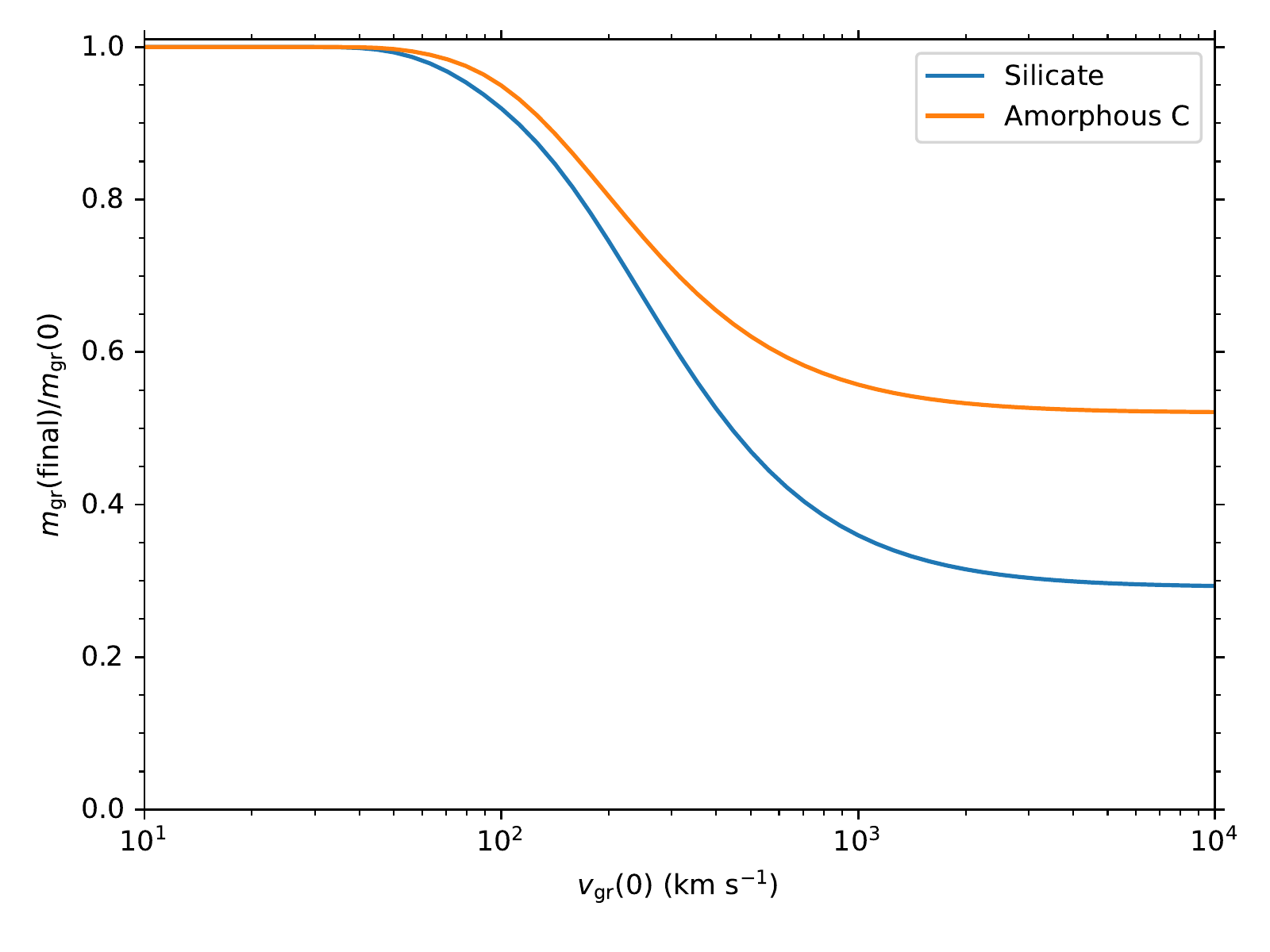}
    \caption{Final surviving grain mass fraction as a function of initial
    grain speed relative to the gas. The assumption here is that only inertial
    sputtering is important and depends on the assumed elemental composition
    of the medium. It is also implicitly assumed that the medium is uniform in
    temperature. The relation plotted allows us to extrapolate to find the
    surviving mass fraction of grains that reach, and are stopped in, the
    uniform ISM surrounding the SN that produced them.}
    \label{fig:mm0_v}
\end{figure}

Once the grains get into the uniform ISM, we can determine analytically how
much more of their mass will be sputtered while they are slowed to a velocity
below the sputtering threshold.  As shown by \citet{Micelotta_etal_2016} this
can be done by combining equations (\ref{eqn:dmdt}) and (\ref{eqn:dvdt}) to
get
\begin{equation}
    {dm_\mathrm{gr}\over dv_\mathrm{gr}} = {m_\mathrm{gr}\over
    v_\mathrm{gr}}\, {m_\mathrm{sp} n_\mathrm{H}\over \rho_\mathrm{gas} \xi}\,
    \sum_i\, A_i\, Y_i(E)
\end{equation}
which can be solved to yield
\begin{equation}
    {m_\mathrm{gr}(t)\over m_\mathrm{gr}(0)} = \exp\left[ {m_\mathrm{sp}
    n_\mathrm{H}\over \rho_\mathrm{gas} \xi}\,
    \int_{v_\mathrm{gr}(0)}^{v_\mathrm{gr}(t)}\, \sum_i A_i\, Y_i(E)
    {dv_\mathrm{gr}\over v_\mathrm{gr}}\right],
\end{equation}
where $E = m_i v_\mathrm{gr}^2/2$. Here we are assuming that the sputtering is
purely inertial since the ISM has a low enough temperature ($T \lesssim 10^4$
K) that thermal sputtering is not important. The ratio of final to initial
mass that results for grains that are slowed below the sputtering threshold
then only depends on the initial grain velocity relative to the gas, the grain
type, and the gas composition. For our assumed grain types and interstellar
gas composition (logarithmic gas phase abundances relative to $\mathrm{H}
= 12$: $A_\mathrm{He} = 11.0$, $A_\mathrm{C} = 8.31$, $A_\mathrm{N} = 7.99$,
$A_\mathrm{O} = 8.83$) we get the curves plotted in Figure \ref{fig:mm0_v}.
Note that the destruction saturates effectively for very high speeds such that
there is a maximum fraction of the initial mass that is sputtered. This
happens because the yield curves turn over at high energy, because the grains
become effectively transparent to the particles for high enough relative
speeds.

\subsubsection{Neglected Processes}
We note that, while we believe that we have included all of the dominant
processes in our calculations, other works have included some that we have
left out. Most importantly we have not included the magnetic field which may
significantly change the grains' trajectories, even possibly returning them to
the remnant after they escape. It is unknown how strong the magnetic field is
in SN ejecta. If it is just the stellar magnetic field diluted via expansion,
it would be extremely small because the stellar dipole field will be diluted
as the cube of the expansion radius. In the ISM it is quite likely that grains
will encounter magnetic fields with typical magnitudes of several microgauss,
which could be important for the trajectories of the grains. However, grains
exiting the remnant at speeds of several hundreds to thousands of kilometers
per second will have gyroradii large enough in most cases that they will not
return to the remnant.

The charge that the grains will have at the low temperatures of the cold or
warm ISM depends on the UV environment and electron density. As described in,
for example \citet{McKee_etal_1987}, the controlling parameter is proportional
to $G_0/n_\mathrm{e}$ where $G_0$ is a measure of the FUV background. The
grain speed also plays an important role. We have found that the gyroradii for
grains escaping the remnant span a wide range from $\sim 1$ pc to hundreds of
parsecs, depending on the parameter values assumed. In addition there is the
question of the morphology of the magnetic field. A uniform field could allow
for reflection of the grains back into the remnant, while a field with a
turbulent component could allow for diffusion of the grains through the ISM.
Recent work by \citet{Fry_etal_2020} finds that a large fraction of grains are
reflected back into the remnant by the magnetic field. However that work
focuses on Fe grains and they find grain charges that differ substantially
from those that we find for the silicate and carbonaceous grains that we study
in this paper. Given the uncertainties in the charging and the interstellar
field we do not consider magnetic field in this work and only note that it
could reduce the grain escape under particular sets of conditions.

Another process that we have neglected is grain-grain collisions. 
Inclusion of grain-grain collisions requires knowledge of the full population
of grains in a region and their relative velocities, which is not feasible
within the confines of our method of following grain trajectories. The results
of grain-grain collisions include shattering, which redistributes grain mass
from large grains to smaller ones, and vaporization, which destroys grains.
For grains that are simply slowing in the ISM, shattering will have little
effect on the total grain mass since smaller grains that are created will slow
quickly and lose a fraction their mass determined by their velocity relative
to the gas. Only vaporization will lead to excess mass loss. This contrasts
with the situation for shocks propagating in the ISM
\citep[e.g.][]{Jones_etal_1996,Slavin_etal_2015}. For fast shocks in the ISM,
$v_\mathrm{shock} \gtrsim 150$ km s$^{-1}$, shattering creates small grains
that are thermally sputtered faster than the equivalent mass in larger grains
during the short period in the post-shock flow that the gas is hot enough for
thermal sputtering to be important. For slower shocks, although grain-grain
collisions dramatically change the grain size distribution, the effect on
grain destruction is minor because the hot post-shock zone is very thin. Thus
we believe that for SN grains that are stopped in the ISM, grain-grain
collisions may change their size distribution but will not affect the total
mass in dust substantially.

For grains propagating in the hot shocked region of the SNR before escaping
into the ISM, grain-grain collisions could occur where larger grains from
deeper in the remnant collide with smaller grains that were slowed closer to
the forward shock. The relatively large volume of the hot shocked region means
that the volume density of grains will be small and the probability of
collisions low.  The high densities in the ejecta clumps and their possibly
low gas-to-dust ratios provide a more likely scenario for grain-grain
collisions being important. Since the grains begin moving outward together, we
do not expect them to have substantial speeds relative to each other in
general. (We assume that there has been no dust formed in the smooth ejecta
because of the much lower density there.) With ejecta clumps that are further
out moving faster initially, we do not see much passing of the farther out
grains by those interior to them. Nevertheless, especially in clumps that
contain a range of grain sizes, it is possible that a significant number of
collisions could occur between slowed small grains and large grains which
would tend to decrease the average grain size and reduce grain survival.
\citet{Kirchschlager_etal_2019} have included grain-grain collisions in their
calculations and find them to have a significant impact on grain destruction,
though their calculations differ from ours in several ways.

A newly proposed process that we also do not include is grains growth via
implantation (or ion trapping) of heavy ions into the grains
\citet{Kirchschlager_etal_2020}. This could increase the grain size in the
ejecta clumps before the grains escape into the ISM. However, since the
initial grain size distribution and the condensation efficiency in the ejecta
is quite uncertain (see \S \ref{sec:results}) one could consider those effects
to be folded into that uncertainty.

Finally, we do not include plasma drag in our calculations. Plasma drag depends on the grain charge and several characteristics of the plasma,
\begin{equation}
    F_d = 4 \pi a^2 k_B T \phi^2 \ln \Lambda \sum_{i} n_i z_i^2 H(s_i),
\end{equation}
where $a$ is the grain size, $T$ is the plasma temperature, $\phi = Z e^2/a
k_B T$ is the potential parameter, $\ln \Lambda$ is the Coulomb logarithm, and
the sum is over elements in the gas phase. See \citet{McKee_etal_1987} and
references therein for details.  We have carried out calculations of the grain
charge, as we did in \cite{Slavin_etal_2015} using methods from
\citet{Weingartner+Draine_2001} with updates from
\citet{Weingartner_etal_2006} and modifications to include the relative
gas-grain velocity. Exploring a variety of conditions like those that the
grains in our simulations are subject to, we find that the plasma drag is very
small compared with the gas drag in all cases.

\section{Results} \label{sec:results}
\begin{figure}[ht!]
    \centering
    \includegraphics[width=\textwidth]{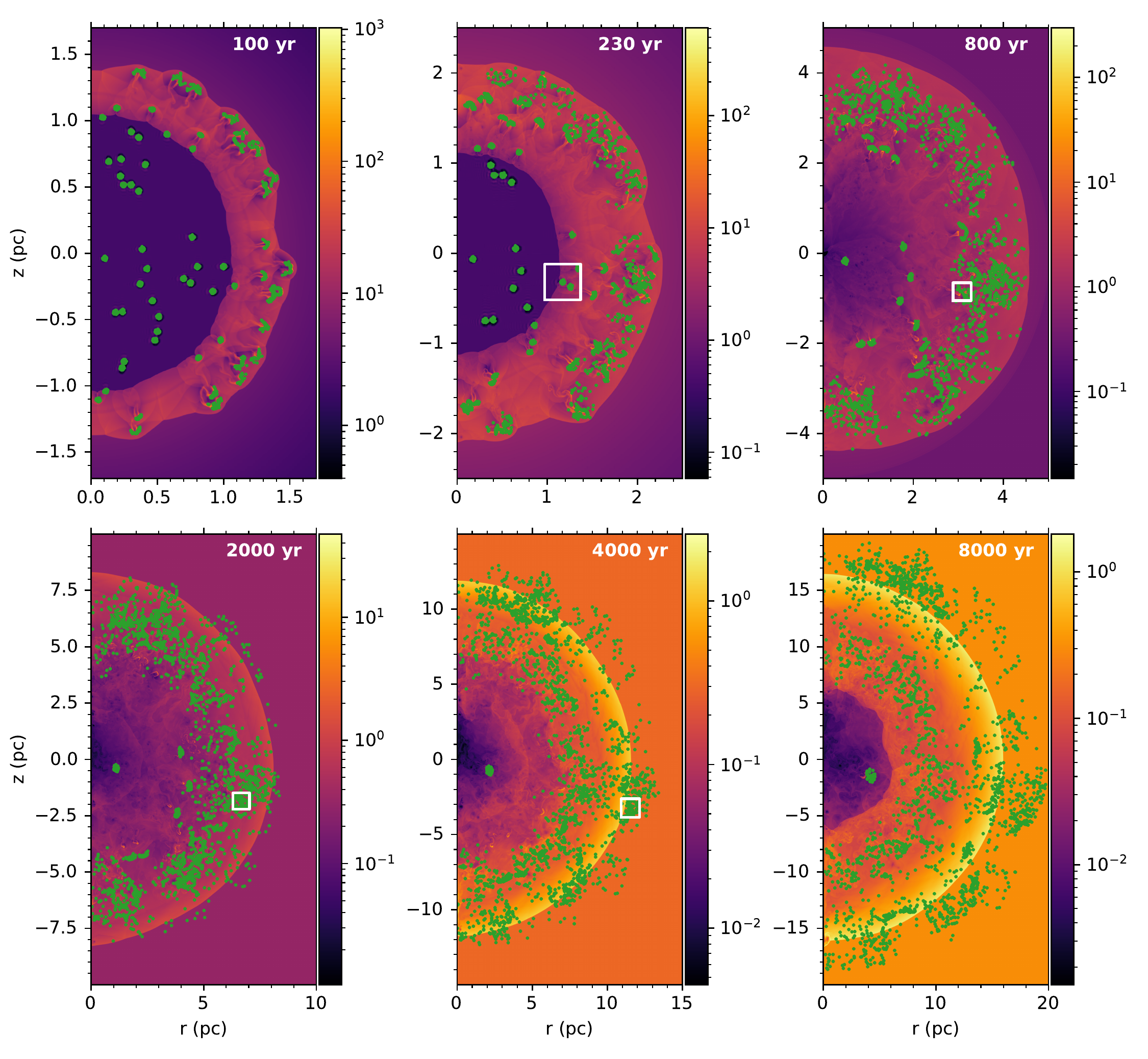}
    \caption{Time evolution of density and particle locations for a simulation
    that includes silicate grains initialized with a radius of 0.1 $\mu$m. The
    background colors show the gas density (in cm$^{-3}$ as indicated in the
    color bars) and the green dots indicate the grain positions. Note that the
    spatial and the density scales vary by as much as an order of magnitude
    between panels. The number of grains is the same in each panel, though at
    early times there is a lot of overlap of grains in clumps, which makes it
    appear that there are fewer grains. The white boxes in some panels show
    the sizes and positions of the panels at the corresponding times in the
    zoomed-in images of Figure \ref{fig:zoomed_image}. (Note however that
    there are more grains in the boxes here because some grains that
    originated in other clumps are included.)}
    \label{fig:image_array}
\end{figure}

Our simulated remnants evolve in the familiar way with a forward shock
forming, a reverse shock propagating backward (relative to the expanding
ejecta), and a contact discontinuity separating the shocked ejecta from the
shocked circumstellar medium. However, because of their high density, the
ejecta clumps decelerate more slowly than the smooth ejecta. Thus after the
reverse shock passes a clump, the shocked smooth ejecta sweeps past it in the
clump frame, leading to a shear flow. This drives turbulence in the shocked
ejecta and the clump, producing a complex velocity field that affects grain
trajectories as the grains escape their natal clumps. The fastest clumps that
start farthest from the explosion center quickly cross the contact
discontinuity and propagate into the shocked circumstellar medium, which
disrupts and smears the contact discontinuity. The reverse shock speed starts
at about 5000 km s$^{-1}$ but then drops to around 2000 km s$^{-1}$, a bit
higher than the observed value of $1000 - 2000$ km s$^{-1}$ derived from
observations of Cas A \citep{Laming+Hwang_2003,Morse_etal_2004}. The
clumpiness of the ejecta lead to variations in the reverse shock speed in both
position and time. The shocks that propagate through the clumps are slowed
because of the jump in density relative to the smooth ejecta, with
$v_s(\mathrm{clump}) \approx v_s(\mathrm{smooth})/\chi^{1/2}$ (under the
assumption of equal ram pressure) with the resulting speeds ranging from $\sim
100 - 350$ km s$^{-1}$. Because of cooling, the shocks in the clumps slow as
they go radiative. The overall evolution of the gas and dust can be seen in
Figure \ref{fig:image_array} where the density is shown (background color) as
well as the dust grain positions.

\begin{figure}[ht!]
    \centering
    \includegraphics[width=\textwidth]{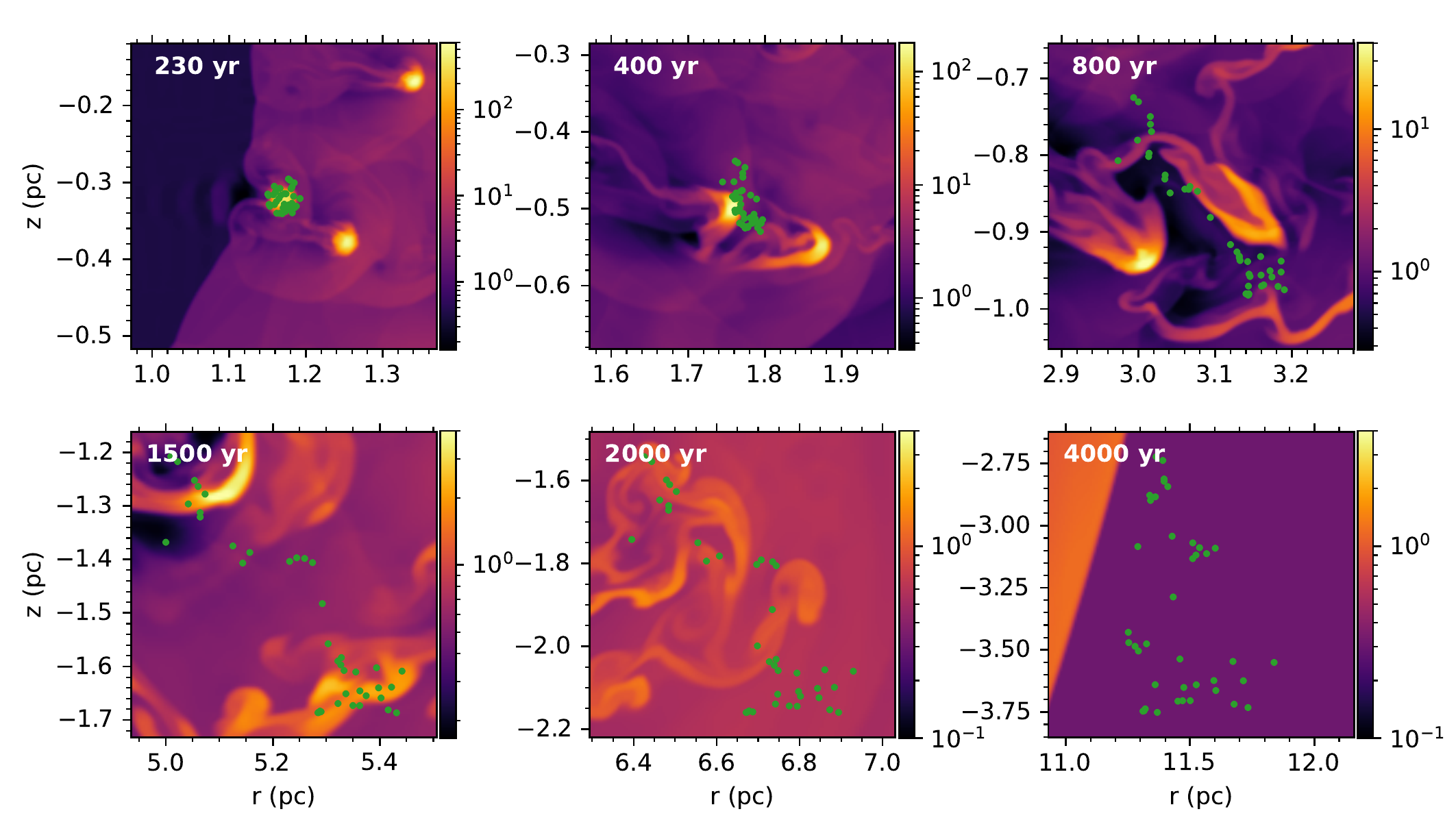}
    \caption{Zoomed in view of the evolution of grains that start in a single
    dense ejecta clump. This is from the same simulation as shown in Figure
    \ref{fig:image_array}. Note the widely divergent paths taken by the grains
    after the clump has been destroyed by the passage of the reverse shock.}
    \label{fig:zoomed_image}
\end{figure}

When the reverse shock encounters the clumps that contain the dust grains, the
clumps are compressed. The shocks that propagate into the clumps are radiative
because of their high density and their high O abundance, which enhances their
cooling. As a result, the clumps cool and are compressed as they are slowed.
The grains then decouple fairly quickly as can be seen in
Figures~\ref{fig:image_array} and especially~\ref{fig:zoomed_image}. The first
few of panels of Figure~\ref{fig:zoomed_image} show how clumps get torn apart
after being shocked. The green dots in that figure show only the grains that
started in a single clump of gas.  In this figure the grains are silicate
grains that have initial radii of 0.1 $\mu$m (as in
Figure~\ref{fig:image_array}). 

\begin{figure}[ht!]
    \centering
    \includegraphics[width=0.9\textwidth]{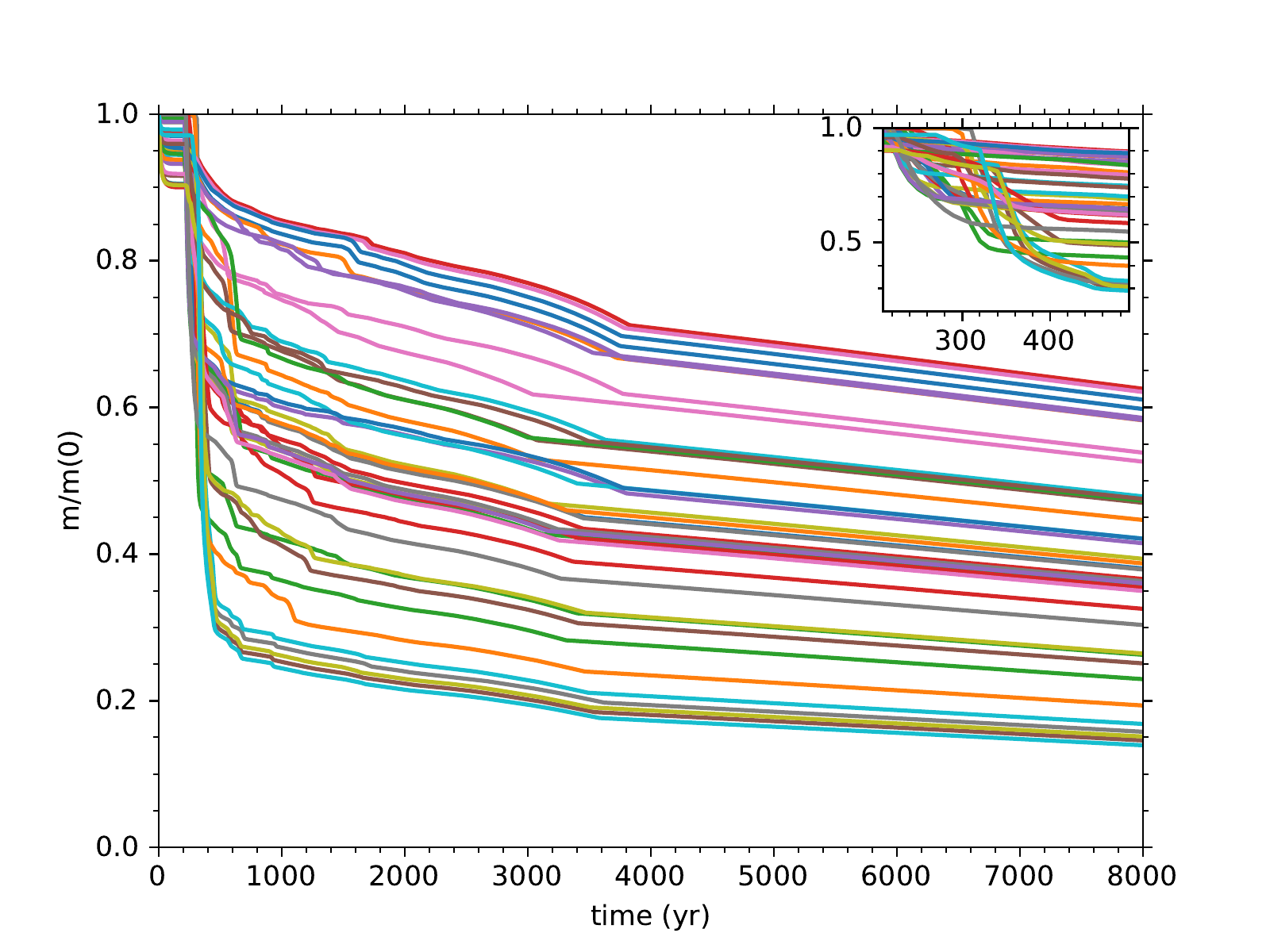}
    \caption{Grain mass evolution for grains in the clump shown in Figure
    \ref{fig:zoomed_image}. The grains lose most of their mass when the
    reverse shock passes through their ejecta clump at $t \sim 200$--300~yr.
    The inset shows a zoomed-in view of the time period when the shock is
    propagating through the clump. The spread in the destruction rates
    reflects the different times of encounters with the reverse shock caused
    by the different locations of the grains in the clump. The grains exit the
    remnant at $t \sim 3800$ yr.  After that they are sputtered more slowly as
    they move through the ISM.}
\label{fig:clump}
\end{figure}

The grains are inertially sputtered on their way out of the clump due to their
speed relative to the gas. The amount of mass lost in traversing a given
column density is proportional to the grain cross section, so larger grains
actually lose more mass than smaller grains, but it is a smaller fraction of
the larger grain's mass. The outer portions of the shocked clumps are hot,
though cooler and denser than the shocked smooth ejecta. This leads to
significant thermal sputtering of the grains in these transition regions.
However, as the grains continue to move out through the remnant, their
destruction rate slows because of the low density of the shock-heated smooth
ejecta. This can be seen in Figure~\ref{fig:clump} where the mass evolution of
each of the grains in Figure \ref{fig:zoomed_image} is shown as a function of
time. The sharp decline in the mass of the grains at $t = 200$--300~yr occurs
as the shock passes through the clump, heating and decelerating it. As can be
seen from the inset, different grains see the effects of the shock at
different times because of their varying locations within the clump. Note that
there is some sputtering before the shock hits because there is some leakage
of grains near the clump edge out of the clump even before the shock hits it,
which leads to motion relative to the clump and inertial sputtering.
After the grains escape their clumps they propagate through the hot shocked
smooth ejecta. While in the hot gas, they suffer thermal sputtering, which can
be seen as the slow mass erosion in Figure~\ref{fig:clump} between $t \sim
500$ yr and $\sim 3800$ yr. The grains exit the remnant at $t \sim 3800$ yr
and after that the sputtering rate decreases and is smoother as the grains are
slowed in the ISM.

\begin{figure}[ht!]
    \centering
    \includegraphics[width=0.7\textwidth]{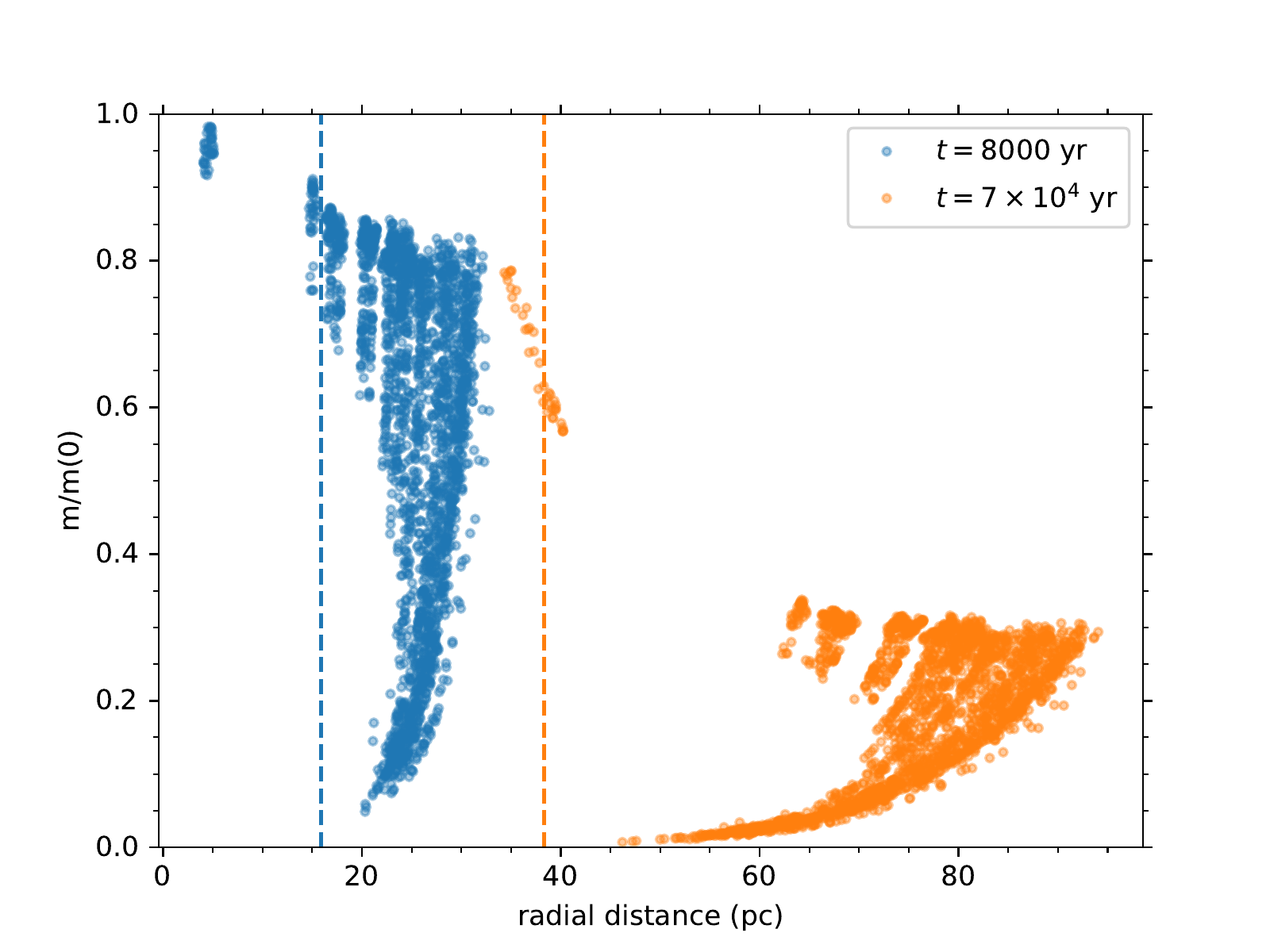}
    \caption{The evolution of grain mass for silicate grains with an initial
    radius of 0.25 $\mu$m. The blue dots show the grain masses vs.\ radial
    distance at the end of the 2D simulation ($t = 8000$ yr). The orange dots
    show the extrapolated evolution to a time $t =7\times 10^4$ yr after the
    explosion when the remnant is beginning to go radiative. The extrapolation
    is done by using the shock radius as calculated by a 1D simulation
    combined with Sedov-Taylor profiles for the density, temperature and
    velocity inside the remnant (see text). In this case nearly all of the
    grains are able to escape the remnant's outer shock (shown as dashed lines
    at each time), though substantial mass loss has occurred between 8000 yr
    and $7\times10^4$ yr.  Further destruction occurs as the grains are slowed
    in the ISM.}
\label{fig:extrap_mass}
\end{figure}

As discussed in \S\ref{sect:extrap}, grain processing does not stop after
escape ahead of the forward shock. The shock speed at the end of our 2D
simulation ($t = 8000$ yr) is 870 km~s$^{-1}$ and its radius is 15.9 pc. Thus
the grains that are already ahead of the shock must have averaged a radial
speed of $\sim 2000$ km s$^{-1}$, though grains still trapped in the remnant
have had lower average speeds.  Indeed we find that even for grains as small
as 0.1 $\mu$m the velocities at that time for grains ahead of the shock range
from $\sim 800$ km s$^{-1}$ to nearly 1400 km s$^{-1}$. Larger grains can have
velocities up to $\sim 4000$ km s$^{-1}$. As can be seen from Figure
\ref{fig:extrap_mass}, this will lead to substantial further loss of mass as
the grains slow in the ISM. Taking 1000 km s$^{-1}$ as an example, a silicate
grain will lose 64\% of its mass as it slows and a carbonaceous grain will
lose 44\% of its mass.

\begin{figure}[ht!]
    \centering
    \includegraphics[width=0.7\textwidth]{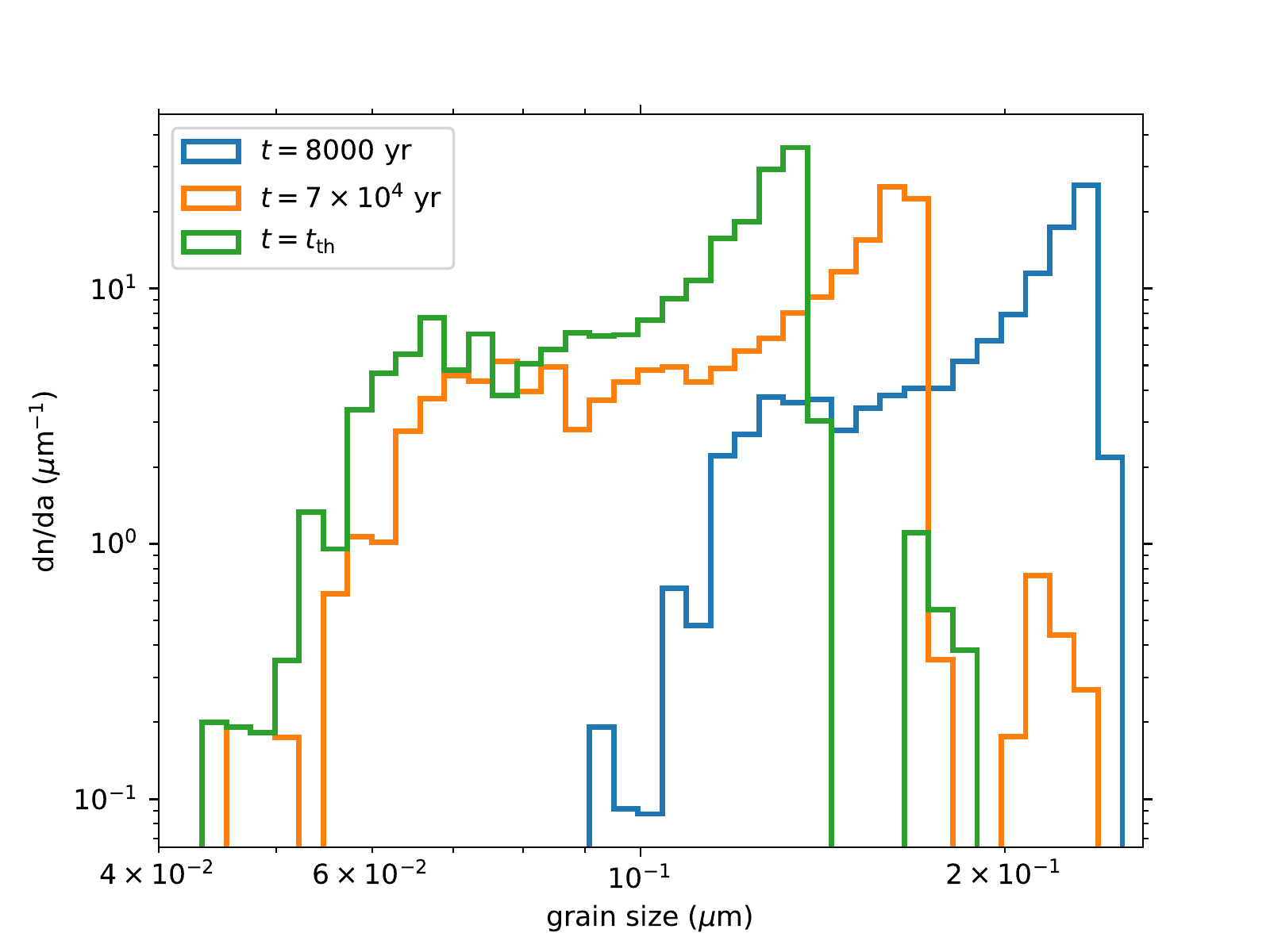}
    \caption{Grain size distributions for the case of silicate grains with
    initial size of 0.25 $\mu$m. The histograms are normalized such that the
    integral over the distribution is 1 and use bins that are constant
    logarithmic intervals. We show the distributions from $t = 8000$ yr (the
    end of the 2D simulation), $t = 7\times10^4$ yr (the shell formation time)
    and $t_\mathrm{th}$ the time when the grains slow to below the sputtering
    threshold. The distribution shapes do not change very much even as the
    total remaining mass in the grains shrinks. Unlike the interstellar grain
    size distribution derived from observations, these are weighted toward the
    largest grain sizes. The actual size distribution injected into the ISM
    will depend on the initial size distribution of the grains formed in the
    ejecta clumps.}
    \label{fig:size_distns}
\end{figure}

Figure \ref{fig:extrap_mass} also illustrates that there is a wide range of
sputtered fractions for different grains depending on where in the remnant and
even where within a clump the grain originated. In Figure
\ref{fig:size_distns} we show the grain size distributions that result for
silicate grains that have initial sizes of 0.25 $\mu$m. Note that all of the
grains in these histograms started with the same grain size.  The size
distribution evolves toward smaller sizes over time while maintaining a
similar shape. Unlike the interstellar grain size distribution, which is
heavily weighted to small grain sizes (by number), these distributions are
weighted to the large grain end. Of course the actual grain size distribution
that is injected into the ISM will depend on the size distribution of the
grains formed. The distributions shown indicate that the final distribution
injected into the ISM will be similar to the initial distribution from grain
formation, though shifted to smaller grain sizes and with an added tail toward
the smaller grain size end.

\begin{deluxetable}{lllcll}
\tablecaption{Grain Mass Survival Fraction \label{tab:survival}}
\tablewidth{0pt}
\tablehead{\colhead{grain type} & \multicolumn2c{silicate} & & 
\multicolumn2c{carbonaceous} \\
\cline{2-3} \cline{5-6}
\colhead{initial size ($\mu$m)} & \colhead{$t_{sh}$} & \colhead{$t_{th}$} & &
\colhead{$t_{sh}$} & \colhead{$t_{th}$}}
%\colhead{time} & \colhead{$t_{sh}$} & \colhead{$t_{th}$} & &
%\colhead{$t_{sh}$} & \colhead{$t_{th}$} \\
%\colhead{initial size ($\mu$m)} & & & & & \\}
\startdata
0.04 & 0.004 & 0.003 & & 0.021 & 0.016 \\
0.1 & 0.028 & 0.022 & & 0.350 & 0.300 \\
0.25 & 0.198 & 0.106 & & 0.702 & 0.460 \\
0.395 & 0.400 & 0.169 & & 0.818 & 0.488 \\
0.625 & 0.602 & 0.218 & & 0.893 & 0.503 \\
\enddata
\tablecomments{$t_{sh} = 7\times10^4$ yr is the shell formation time for the remnant. $t_{th}$ is the time at which the grain has slowed below the threshold speed for sputtering, $v_{th} \sim 20$ km s$^{-1}$, which varies from grain to grain.}
\end{deluxetable}

We list the the final results of our calculations of the fraction of initial
mass remaining for grains of different types and initial sizes in Table
\ref{tab:survival}. We give both the mass fraction of grains remaining at the
end of the 1D simulation extrapolation (essentially the shell formation time
for the remnant, $t_\mathrm{sh} = 7\times10^4$ yr) and at the time
$t_\mathrm{th}$ that each grain is slowed to the threshold velocity
$v_\mathrm{thresh}$ for inertial sputtering. The threshold velocity for both
silicate and carbonaceous grains is $\sim 20$ km s$^{-1}$. It is clear that
for the largest grain sizes the fraction of mass remaining is almost as much
as would be expected if the grains had been directly released into the ISM at
high speeds. Thus if even larger grains are formed in SNe, we expect their
surviving mass fraction to be roughly the same as for the largest grains in
the table.

\begin{deluxetable}{cDD}
\tablecaption{Overall Grain Mass Survival Fractions for Different Initial Size Distributions\label{tab:size_distn}}
\tablewidth{0pt}
\decimals

% mass fraction results:
% Silicate, log-normal dist, apeak = 0.1:  0.0316
% Silicate, log-normal dist, apeak = 0.25:  0.1281
% Silicate, log-normal dist, apeak = 0.5:  0.2077

% Carbonaceuos, log-normal dist, apeak = 0.1:  0.3478
% Carbonaceous, log-normal dist, apeak = 0.25:  0.4698
% Carbonaceous, log-normal dist, apeak = 0.5:  0.5008
%
% Silicate, power law, amin = 0.005, amax = 0.25, pl = -3.5:  0.0405
% Silicate, power law, amin = 0.05, amax = 1.0, pl = -3.5:  0.1385
% Silicate, power law, amin = 0.04, amax = 1.0, pl = -4.4:  0.0702
%
% Carbonaceous, power law, amin = 0.005, amax = 0.25, pl = -3.5:  0.3162
% Carbonaceous, power law, amin = 0.05, amax = 1.0, pl = -3.5:  0.4390
% Carbonaceous, power law, amin = 0.04, amax = 1.0, pl = -4.4:  0.3166

\tablehead{\colhead{size distribution\tablenotemark{a}} & \multicolumn2c{silicate} & 
\multicolumn2c{carbonaceous}}
\startdata
LN1 & 0.0316 & 0.348 \\
LN2 & 0.128 & 0.470 \\
LN3 & 0.208 & 0.501 \\
PL1 & 0.0405 & 0.316 \\
PL2 & 0.139 & 0.439 \\
PL3 & 0.0702 & 0.317 \\
\enddata
\tablenotetext{a}{The grain size distributions are described in the text %ASH add
and shown in Figure~\ref{fig:size_distns} (solid lines).
LN\# are log-normal distributions and PL\# are power-law distributions.}
\end{deluxetable}

To determine how much dust SNe can inject into the ISM, we need a few more
pieces of information: the total mass that is condensed into grains and the
grain size distribution for the grains formed. The mass of metals produced by
SNe has been estimated in a number of studies. As an example,
\citet{Woosley+Heger_2007} have calculated the mass of metals in the ejecta of
the explosion of a star with initial mass 19 M$_\sun$. Using their results, we
calculate the maximum dust masses in the ejecta assuming that all the Mg, Si,
Fe and 4/3 of their combined number of O atoms were locked up in silicate dust
of the form MgSiFeO$_4$ and that all the C was locked up in carbon dust. Doing
this we find that the SN can produce 0.473~M$_\sun$ of silicate grains and
0.209~M$_\sun$ of carbonaceous grains.

\begin{figure}[ht!]
    \centering
    \includegraphics[width=0.7\textwidth]{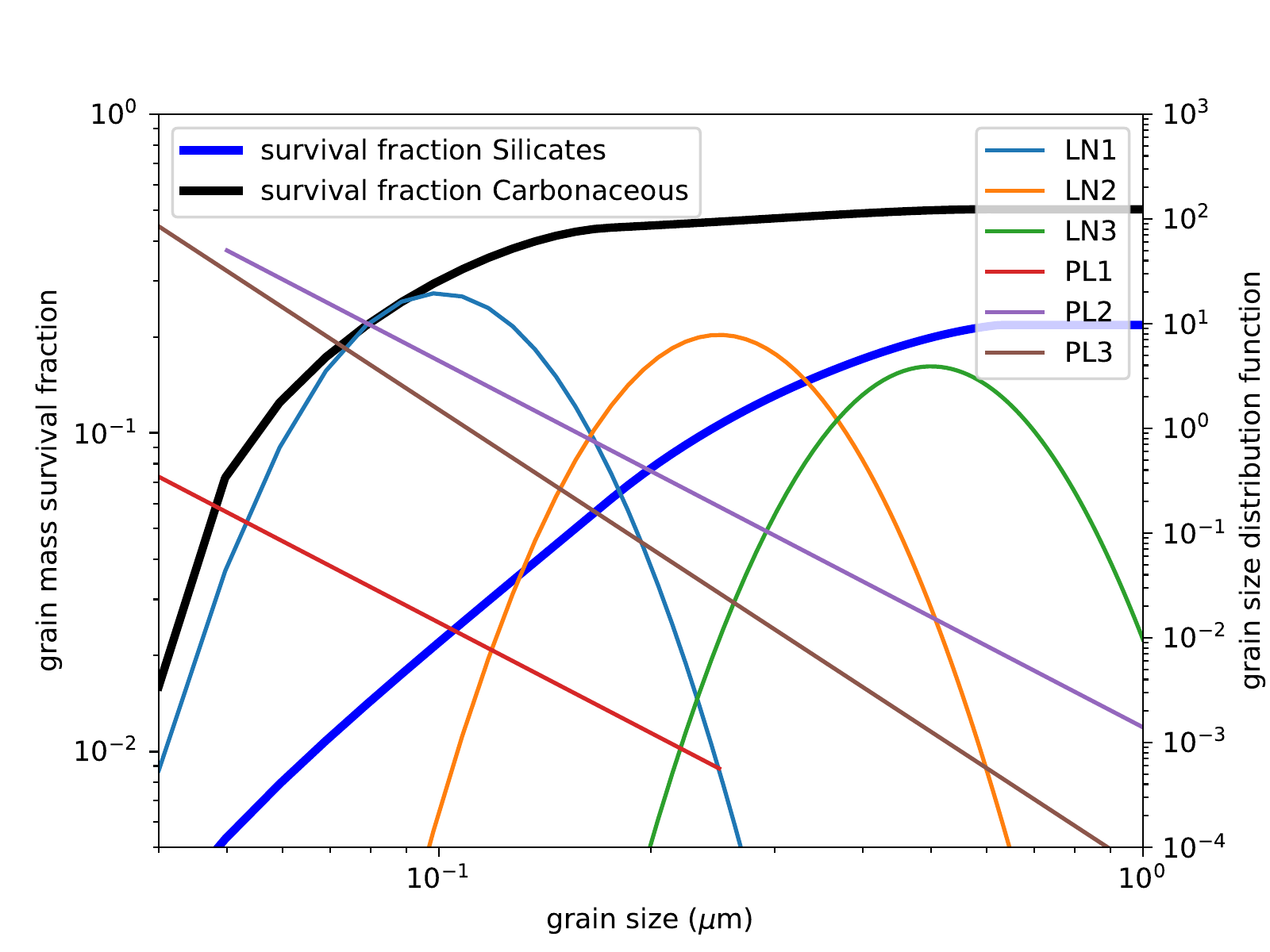}
    \caption{Dust mass survival fractions for silicate and carbonaceous grains
    (thick lines and left axis) and the grain size distributions we have
    examined (thin lines, right axis). The survival fractions are just those
    in Table \ref{tab:survival} (each point calculated for a single grain
    size) with a quadratic spline interpolation. The parameters for the size
    distributions are detailed in the text. Note that for PL1 and PL2 the
    power laws extend to smaller sizes than shown in the plot. Such small
    grains are all destroyed inside the remnant, but are important to include
    for the overall normalization. The overall survival fraction is the
    survival fraction as a function of mass integrated over the initial mass
    distribution of the formed grains (i.e. the size distribution times the
    grain mass).}
    \label{fig:surv_size_distn}
\end{figure}

As for grain size distributions, the most comprehensive calculations of grain
formation in the dense ejecta of SNe are from \citet{Sarangi+Cherchneff_2015}
and \citet{Sluder_etal_2018}. Neither work characterizes the size distribution
in a simple way and such calculations are difficult and subject to a variety
of uncertainties. Rather than attempt to use their results then, we have
chosen to use a couple of simple characterizations, namely a power law,
\begin{equation}
    \frac{dn}{da} = \frac{1 + \gamma}{(a_\mathrm{max}^{1 + \gamma} -
    a_\mathrm{min}^{1 + \gamma})}\, a^{-\gamma},
\end{equation}
and a log-normal distribution,
\begin{equation}
    \frac{dn}{da} = \frac{1}{a \sigma \sqrt{2 \pi}}
    \exp\left(\frac{-(\ln(a/a_\mathrm{peak}) - \sigma^2)^2}{2
    \sigma^2}\right),
\end{equation}
where the distributions are normalized such that the integral of $dn/da$ over
the full size range is equal to unity. For the log-normal distribution the
parameters are the grain size at the peak of the distribution,
$a_\mathrm{peak}$ and the width parameter, $\sigma$. We have assumed that
$\sigma = 0.2$ in all cases. For the power law distributions, we need to
assume the minimum and maximum grain sizes, $a_\mathrm{max}$ and
$a_\mathrm{min}$, as well as the value of $\gamma$.  The size distributions
that we have examined are as follows:
\begin{itemize}
    \item LN1: log-normal with peak at 0.1 $\mu$m
    \item LN2: log-normal with peak at 0.25 $\mu$m
    \item LN3: log-normal with peak at 0.5 $\mu$m
    \item PL1: \citet{Mathis_etal_1977} style power law, $a_\mathrm{min} =
        0.005\,\mu$m, $a_\mathrm{max} = 0.25\,\mu$m, $\gamma = 3.5$
    \item PL2: $a_\mathrm{min} = 0.05\,\mu$m, $a_\mathrm{max} = 1.0\,\mu$m,
        $\gamma = 3.5$
    \item PL3: $a_\mathrm{min} = 0.04\,\mu$m, $a_\mathrm{max} = 1.0\,\mu$m,
        $\gamma = 4.4$
\end{itemize}
These size distributions are illustrated in Figure \ref{fig:surv_size_distn}
where the survival fractions from Table \ref{tab:survival}, interpolated by a
quadratic spline, are also plotted. PL3 has the power law exponent that
\citet{Sluder_etal_2018} find for their results, though the $a_\mathrm{min}$
and $a_\mathrm{max}$ values we have chosen do not match their results closely. 

To calculate the overall fraction of grains that survive and are injected into
the ISM, we need to integrate the mass distribution of grains over the grain
mass survival fraction. The mass distribution is just the size distribution
times the mass of a grain, $(4/3) \pi a^3 \rho_\mathrm{gr}$, with proper
normalization, where $\rho_\mathrm{gr}$ is the mass density of the solid grain
material. We have carried out such calculations for the size distributions and
the results are listed in Table \ref{tab:size_distn}. From these it is clear
that any size distribution that has a large grain cutoff that is too small
results in a small fraction of silicates grains surviving. Only the LN3
distribution leads to more than 20\% survival. On the other hand, the
relatively large fraction of even small carbonaceous grains that survive
results in an overall survival fraction that is not highly sensitive to our
assumed size distribution, with more than 30\% survival in all cases and about
50\% survival for distribution LN3. This is primarily due to the lower
sputtering yield for carbonaceous grains, especially near the peak of the
yield at $v \sim 500 - 1000$ km s$^{-1}$.

The total effective mass of dust injected into the ISM is the product of the
overall survival fraction, as listed in Table \ref{tab:size_distn}, and the
total mass of dust (of the particular type) produced by the SN. If we take the
values of 0.473 M$_\sun$ for silicate dust and 0.209 M$_\sun$ for carbonaceous
dust as the maximum masses of dust that can be produced by a 19 $M_\sun$ star,
as referenced above, then the total mass of dust injected into the ISM is
0.098 $M_\sun$ of silicate dust and 0.105 $M_\sun$ of carbonaceous dust per
SN. Here we have assumed the LN3 size distribution. Based on our calculations,
this is essentially the maximum amount of dust that can be injected since this
size distribution weights large grains that have close to a maximal mass
survival fraction. Note that these values also assume 100\% conversion of
grain constituent elements in the SN into grains. Near total conversion into
grains is supported by observations of SN 1987A \citep{Matsuura_etal_2015} and
Cas A \citep{Barlow_etal_2010,Arendt_etal_2014}. It is clear that less
optimistic assumptions can lead to much smaller injection rates for silicate
grains in particular. For carbonaceous grains, even a size distribution
weighted toward smaller grains such as PL3 still allows for as much as
0.07~M$_\sun$ of grain injection.

\section{Conclusions}
We have shown that dust grains formed in the dense ejecta of SNe can escape
from the clumps and survive the reverse shock. The grains decouple from the
gas and stream outward in the remnant. Grains that are large enough, $\sim
0.25\,\mu$m for silicates and $\sim 0.1\,\mu$m for carbonaceous grains, can
escape ahead of the forward shock into the surrounding ISM. However, the
grains must slow after reaching the ISM to be considered as contributing to
the dust content of the medium, which entails further erosion via inertial
sputtering. Including the sputtering while slowing sets the upper limit on the
survival mass fraction of SN-created grains. For an initial grain mass
distribution weighted toward large, $a \sim 0.5\,\mu$m, grains, roughly 20\%
(by mass) of silicate grains and up to 50\% of carbonaceous grains are
injected into the ISM. Considering the case of a 19 M$_\sun$ star, the amount
of mass in the grain constituents in the SN leads us to expect at most about
0.1 M$_\sun$ in either silicate or carbonaceous grains can survive injection
into the ISM.

\acknowledgments
This work has been supported by NASA Theory Program grant No.\ NNX17AH80G. We
have benefited from the use of high performance computing resources of NASA's
Pleiades cluster as well as the Smithsonian Institution's Hydra cluster. The
code FLASH used in this work was in part developed by the DOE NNSA-ASC OASCR
Flash Center at the University of Chicago. We thank Elisabetta Micelotta for
helpful conversations.

\vspace{5mm}
\facilities{Smithsonian's Hydra cluster, NASA's Pleiades HPC cluster}


% \software{astropy \citep{2013A&A...558A..33A}}
% mm [added missing citations]
\software{FLASH
    \citep[\url{http://flash.uchicago.edu/site/flashcode/}]{Fryxell_etal_2000,
    Dubey_etal_2012}, python \citep{VanRossum_1995}, numpy
    \citep{Oliphant_2006}, scipy \citep{Jones_etal_2001}, yt
    \citep{Turk_etal_2011}, Cloudy \citep{Ferland_etal_2017}}



\bibliography{CasA_dust}{}
\bibliographystyle{aasjournal}

\end{document}